\documentclass[12pt,a4paper,onecolumn]{article}
\usepackage{epsfig}
\begin{document}
\title{The star trellis decoding of Reed-Solomon codes}
\author{Sergei V. Fedorenko \\
Department of safety in information systems \\
St.Petersburg State University of Aerospace Instrumentation \\
190000, Bolshaya Morskaia, 67, St.Petersburg, Russia}

\maketitle

\begin{abstract}
The new method for Reed-Solomon codes decoding is introduced.
The method is based on the star trellis decoding of the binary image
of Reed-Solomon codes.
\end{abstract}

\section{The Golay code in star representation}

For the Golay code ${\cal G}_{Golay}$ with codelength $n=24$ the star trellis 
was proposed in \cite{SF}.

The time axis consists of a number of parts $I=\cup_j I_j$, which are joint in
one common point.
In Figure 1 an example factor graph representation is
depicted. This example has three equal length parts.
The time axis of the parts are given by
$I_j=\{0,(j-1) n/3 +1, \ldots,j n/3,\infty\}$,
where $\infty$ denotes the junction. 
This junction in Figure 1 is represented by a square 
and the state space is denoted $S_i$.

\begin{figure}[ht]
\begin{center}
\begin{picture}(0,0)%
\epsfig{file=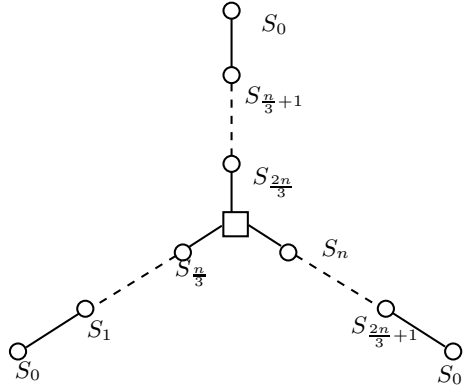}%
\end{picture}%
\setlength{\unitlength}{3355sp}%
\begingroup\makeatletter\ifx\SetFigFont\undefined%
\gdef\SetFigFont#1#2#3#4#5{%
  \reset@font\fontsize{#1}{#2pt}%
  \fontfamily{#3}\fontseries{#4}\fontshape{#5}%
  \selectfont}%
\fi\endgroup%
\begin{picture}(3486,2806)(558,-2041)
\put(3835,-2041){\makebox(0,0)[b]{\smash{\SetFigFont{9}{10.8}{\rmdefault}{\mddefault}{\updefault}$S_0$}}}
\put(3360,-1684){\makebox(0,0)[b]{\smash{\SetFigFont{9}{10.8}{\rmdefault}{\mddefault}{\updefault}$S_{\frac{2n}{3}+1}$}}}
\put(2551, 14){\makebox(0,0)[b]{\smash{\SetFigFont{9}{10.8}{\rmdefault}{\mddefault}{\updefault}$S_{\frac{n}{3}+1}$}}}
\put(2551,539){\makebox(0,0)[b]{\smash{\SetFigFont{9}{10.8}{\rmdefault}{\mddefault}{\updefault}$S_0$}}}
\put(2551,-586){\makebox(0,0)[b]{\smash{\SetFigFont{9}{10.8}{\rmdefault}{\mddefault}{\updefault}$S_{\frac{2n}{3}}$}}}
\put(1951,-1261){\makebox(0,0)[b]{\smash{\SetFigFont{9}{10.8}{\rmdefault}{\mddefault}{\updefault}$S_{\frac{n}{3}}$}}}
\put(1276,-1711){\makebox(0,0)[b]{\smash{\SetFigFont{9}{10.8}{\rmdefault}{\mddefault}{\updefault}$S_1$}}}
\put(751,-2011){\makebox(0,0)[b]{\smash{\SetFigFont{9}{10.8}{\rmdefault}{\mddefault}{\updefault}$S_0$}}}
\put(3001,-1111){\makebox(0,0)[b]{\smash{\SetFigFont{9}{10.8}{\rmdefault}{\mddefault}{\updefault}$S_n$}}}
\end{picture}
\caption{Factor Graph Representation of the Star Trellis.}
\end{center}
\end{figure}           

Each part of the time axis is associated with the conventional 
trellis shortening 
with a single starting state and some end states. 
The star trellis consists of a union of all conventional trellis shortening 
in the junction $\infty$. The star trellis of the Golay code is given 
in Figure 2.

\begin{figure}[ht]
\begin{center}
\begin{picture}(0,0)%
\epsfig{file=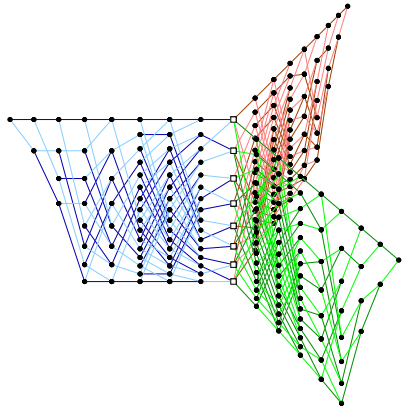}%
\end{picture}%
\setlength{\unitlength}{1657sp}%
\begingroup\makeatletter\ifx\SetFigFont\undefined%
\gdef\SetFigFont#1#2#3#4#5{%
  \reset@font\fontsize{#1}{#2pt}%
  \fontfamily{#3}\fontseries{#4}\fontshape{#5}%
  \selectfont}%
\fi\endgroup%
\begin{picture}(5826,5998)(718,-6432)
\end{picture}
\caption{The Star Trellis of the Golay Code.}
\end{center}
\end{figure}           

${\cal G}_{Golay}$ may be represented by the following generator matrix:
$$
G_{Golay} = 
\left[ 
\begin{array}{c|c|c}
 1 1 1 1 0 0 0 0 & 0 0 0 0 0 0 0 0 & 1 1 1 1 0 0 0 0 \cr
 0 1 0 1 1 0 1 0 & 0 0 0 0 0 0 0 0 & 0 1 0 1 1 0 1 0 \cr
 0 0 1 1 1 1 0 0 & 0 0 0 0 0 0 0 0 & 0 0 1 1 1 1 0 0 \cr\hline 
 0 0 0 0 0 0 0 0 & 1 1 1 1 0 0 0 0 & 1 1 1 1 0 0 0 0 \cr
 0 0 0 0 0 0 0 0 & 0 1 0 1 1 0 1 0 & 0 1 0 1 1 0 1 0 \cr
 0 0 0 0 0 0 0 0 & 0 0 1 1 1 1 0 0 & 0 0 1 1 1 1 0 0 \cr\hline 
 1 1 1 1 1 1 1 1 & 0 0 0 0 0 0 0 0 & 0 0 0 0 0 0 0 0 \cr
 0 0 0 0 0 0 0 0 & 1 1 1 1 1 1 1 1 & 0 0 0 0 0 0 0 0 \cr
 0 0 0 0 0 0 0 0 & 0 0 0 0 0 0 0 0 & 1 1 1 1 1 1 1 1 \cr\hline 
 1 0 0 1 1 0 1 0 & 1 0 0 1 1 0 1 0 & 1 0 0 1 1 0 1 0 \cr
 1 1 0 0 1 0 0 1 & 1 1 0 0 1 0 0 1 & 1 1 0 0 1 0 0 1 \cr
 0 1 1 1 1 0 0 0 & 0 1 1 1 1 0 0 0 & 0 1 1 1 1 0 0 0 \cr
\end{array}
\right].
$$
\noindent
This result was obtained by a permutation of the {\em Turyn}-construction
\cite[18.7.4]{MWS}. 

Each trellis shortening consists of $n/3=8$ sections and has 
a single starting state and eight possible end states. 
The end states correspond to the three last rows of 
the generator matrix $G_{Golay}$. 
We connect three corresponding end-states in eight special states 
to obtain a star trellis for the Golay code.

A valid Golay codeword is one-to-one correspondence union of 
three paths on trellis shortening, which starts in a single starting state 
and ends in one of the eight possible end states. 
For all eight possibilities the special linear dependencies 
need to be satisfied in order to obtain a valid Golay codeword.

\section{The Vardy--Be'ery decomposition}

The map of the Reed-Solomon (RS) code ${\cal RS}$ into its binary image 
$\hbox{Im}({\cal RS})$ was presented in \cite{VB}.

Let us introduce some notations for the RS code.
Let ${\cal RS}$ be a $(N,K,D)$ RS code of length $N=2^m-1$, 
dimension $K$ and minimum Hamming distance $D=N-K+1$ 
over $GF(2^m)$. 
The RS code generator polynomial is $G(x)$
with roots $\alpha, \alpha^2, \ldots , \alpha^{D-1}$,
where $\alpha$ is a primitive element of $GF(2^m)$. 

By analogy, let us introduce some notations for the 
Bose-Chaudhuri-Hocquenghem (BCH) code ${\cal BCH}$ 
with the same parity-check matrix that the ${\cal RS}$ has.
Let ${\cal BCH}$ be a binary $(n,k,d)$ BCH code of length $n=N$, 
dimension $k \le K$ and minimum Hamming distance $d \ge D$.
The BCH code generator polynomial is $g(x) \in GF(2)[x]$
with roots $\alpha, \alpha^2, \ldots , \alpha^{D-1}$
and their cyclotomic conjugates over GF(2).
It is obvious that $G(x) \mid g(x)$.
The ${\cal BCH}$ code has a generator matrix $G_{BCH}$.

Let $\{ \gamma_1, \gamma_2, \ldots , \gamma_m \}$  be any basis in $GF(2^m)$.

For any element $\alpha^j = \sum_{i=1}^{m} a_i \gamma_j \in GF(2^m)$
let us introduce its binary image 
$\hbox{Im}(\alpha^j) = (a_1,a_2,\ldots,a_m)$;
and the binary image of 0 is $\hbox{Im}(0) = (0,0,\ldots,0)$.

Without loss of generality, we shall use a standard basis \\
$\{ \alpha^0, \alpha^1, \alpha^2 , \ldots , \alpha^{m-1} \}$.

For any codeword of the ${\cal RS}$ code we have

$c = (c_0,c_1,\ldots,c_{n-1}) \in {\cal RS}$,

$\gamma_i c = (\gamma_i c_0, \gamma_i c_1,\ldots,\gamma_i c_{n-1}) \in {\cal RS}, \, i \in [1,m]$,

$$
\left[ 
\begin{array}{c|c|c|c}
\gamma_1 c_0 & \gamma_1 c_1 & \ldots & \gamma_1 c_{n-1} \cr\hline 
\gamma_2 c_0 & \gamma_2 c_1 & \ldots & \gamma_2 c_{n-1} \cr\hline 
\ldots       & \ldots       & \ldots & \ldots           \cr\hline 
\gamma_m c_0 & \gamma_m c_1 & \ldots & \gamma_m c_{n-1} \cr
\end{array}
\right] \in {\cal RS},
$$

$$
\left[ 
\begin{array}{c|c|c|c}
\hbox{Im}(\gamma_1 c_0) & \hbox{Im}(\gamma_1 c_1) & \ldots & \hbox{Im}(\gamma_1 c_{n-1}) \cr\hline 
\hbox{Im}(\gamma_2 c_0) & \hbox{Im}(\gamma_2 c_1) & \ldots & \hbox{Im}(\gamma_2 c_{n-1}) \cr\hline 
\ldots                  & \ldots                  & \ldots & \ldots                      \cr\hline 
\hbox{Im}(\gamma_m c_0) & \hbox{Im}(\gamma_m c_1) & \ldots & \hbox{Im}(\gamma_m c_{n-1}) \cr
\end{array}
\right] \in \hbox{Im}({\cal RS}).
$$

Any codeword of the ${\cal BCH}$ code is also a codeword of the ${\cal RS}$ code.

We have
$b = (b_0,b_1,\ldots,b_{n-1}) \in {\cal BCH}, \, b_i \in GF(2)$,

$b = (b_0,b_1,\ldots,b_{n-1}) \in {\cal RS}$,

$\gamma_i b = (\gamma_i b_0, \gamma_i b_1,\ldots,\gamma_i b_{n-1}) \in {\cal RS}, \, i \in [1,m]$.

We use the standard basis $\gamma_i = \alpha^{i-1}$,  \, $i \in [1,m]$,
and obtain 
$$
\begin{array}{ccccccccc}
                                                          & 0 & \ldots & i-1 & i   & i+1 & \ldots & m-1 &    \cr
\hbox{Im}(\gamma_{i+1} b_j) = \hbox{Im}(\alpha^i b_j) = ( & 0 & \ldots & 0   & b_j & 0   & \ldots & 0   & ), \cr
\end{array}
$$

$\gamma_{i+1} b = (\alpha^i b_0, \alpha^i b_1,\ldots,\alpha^i b_{n-1}) \in {\cal RS}, \, i \in [0,m-1]$,

$$
I_b = 
\left[ 
\begin{array}{c|c|c|c}
b_0 0 \ldots 0 & b_1 0 \ldots 0  & \ldots & b_{n-1} 0 \ldots 0  \cr\hline 
0 b_0 \ldots 0 & 0 b_1 \ldots 0  & \ldots & 0 b_{n-1} \ldots 0  \cr\hline 
\ldots         & \ldots          & \ldots & \ldots              \cr\hline 
0 0 \ldots b_0 & 0 0  \ldots b_1 & \ldots & 0 0  \ldots b_{n-1} \cr
\end{array}
\right] \in \hbox{Im}({\cal RS}).
$$

Let us introduce a permutation for the columns of the matrix $I_b$:

$\hbox{Per}\Big((0,0),(0,1),\ldots,(0,m-1) \mid (1,0),(1,1),\ldots,(1,m-1) \mid 
\ldots \mid \\ (n-1,0),(n-1,1),\ldots,(n-1,m-1)\Big)
= \\ \Big((0,0),(1,0),\ldots,(n-1,0) \mid (0,1),(1,1),\ldots,(n-1,1) \mid 
\ldots \mid (0,m-1), \\ (1,m-1),\ldots,(n-1,m-1)\Big)$.

Thus,
$$
\hbox{Per}(I_b) = 
\left[ 
\begin{array}{c|c|c|c}
b_0 b_1 \ldots b_{n-1} & 0 0 \ldots 0           & \ldots & 0 0 \ldots 0           \cr\hline 
0 0 \ldots 0           & b_0 b_1 \ldots b_{n-1} & \ldots & 0 0 \ldots 0           \cr\hline 
\ldots                 & \ldots                 & \ldots & \ldots                 \cr\hline 
0 0 \ldots 0           & 0 0  \ldots 0          & \ldots & b_0 b_1 \ldots b_{n-1} \cr
\end{array}
\right] \in \hbox{Per}(\hbox{Im}({\cal RS})).
$$

It is correct for any codeword $b \in {\cal BCH}$.
Hence the generator matrix of the permutation 
for the binary image RS code may be represented as
$$
{\hbox{\Large{$G$}}}_{\hbox{Per}\left(\hbox{Im}({\cal RS})\right)} = 
\left[ 
\begin{tabular}{c|c|c|c}                       
$G_{BCH}$ & 0       & \ldots & 0         \\ \hline      
0       & $G_{BCH}$ & \ldots & 0         \\ \hline 
\ldots  & \ldots    & \ldots & \ldots    \\ \hline 
0       & 0         & \ldots & $G_{BCH}$ \\ \hline 
\multicolumn{4}{c}{glue vectors}\\
\end{tabular}
\right],
$$
where the submatrix ``glue vectors'' is a $m(K-k) \times Nm$ matrix.

We consider the ${\cal RS}$ (7,5,3) code. 
The permutation of columns of the generator matrix for the binary image RS code  
is the generator matrix for the binary (21,15,3) code \cite{HPGC}:
$$
{\hbox{\Large{$G$}}}_{\hbox{Per}\left(\hbox{Im}({\cal RS})\right)} = 
\left[ 
\begin{array}{c|c|c}
 1 1 0 1 0 0 0 & 0 0 0 0 0 0 0 & 0 0 0 0 0 0 0 \cr
 0 1 1 0 1 0 0 & 0 0 0 0 0 0 0 & 0 0 0 0 0 0 0 \cr
 0 0 1 1 0 1 0 & 0 0 0 0 0 0 0 & 0 0 0 0 0 0 0 \cr
 0 0 0 1 1 0 1 & 0 0 0 0 0 0 0 & 0 0 0 0 0 0 0 \cr\hline 
 0 0 0 0 0 0 0 & 1 1 0 1 0 0 0 & 0 0 0 0 0 0 0 \cr
 0 0 0 0 0 0 0 & 0 1 1 0 1 0 0 & 0 0 0 0 0 0 0 \cr
 0 0 0 0 0 0 0 & 0 0 1 1 0 1 0 & 0 0 0 0 0 0 0 \cr
 0 0 0 0 0 0 0 & 0 0 0 1 1 0 1 & 0 0 0 0 0 0 0 \cr\hline 
 0 0 0 0 0 0 0 & 0 0 0 0 0 0 0 & 1 1 0 1 0 0 0 \cr
 0 0 0 0 0 0 0 & 0 0 0 0 0 0 0 & 0 1 1 0 1 0 0 \cr
 0 0 0 0 0 0 0 & 0 0 0 0 0 0 0 & 0 0 1 1 0 1 0 \cr
 0 0 0 0 0 0 0 & 0 0 0 0 0 0 0 & 0 0 0 1 1 0 1 \cr\hline 
 1 0 0 0 0 0 0 & 0 0 0 0 1 0 0 & 0 0 1 0 0 0 0 \cr
 0 1 0 0 0 0 0 & 0 0 0 0 0 1 0 & 0 0 0 1 0 0 0 \cr
 0 0 1 0 0 0 0 & 0 0 0 0 0 0 1 & 0 0 0 0 1 0 0 \cr
\end{array}                                
\right].
$$                                         

\section{Decoding method}

The star trellis can be constructed for any Reed-Solomon code.
The star trellis consists of $m$ parts.
Each part is a conventional trellis shortening 
with a single starting state and $2^{m(K-k)}$ end states. 
The end states are defined by ``glue vectors'' in the
generator matrix of the permutation for the binary image RS code.

The decoding method has two stages.
The first stage is the soft-decision decoding for $m$ trellis shortening.
The result of this stage is a list of codewords.
The cardinality of the list is not more than $2^{m(K-k)}$.
On the second stage the nearest codeword form the list
to the received vector is chosen.

The simulation for the ${\cal RS}$ (7,5,3) code is executed. 
Bit Error Rate (BER) and Codeword Error Rate (CER) 
performance dependence on signal-to-noise ratio (SNR) for 
additive white Gaussian noise (AWGN) channel is given in Table 1.
By classic decoding method for RS code we understand 
decoding of both errors and erasures 
(see, for example, Forney or Chase algorithms \cite{F, C})
with the Berlekamp-Massey algorithm for the key equation solving.

\bigskip
\rightline{Table 1}
\nopagebreak
\medskip
\centerline{
\begin{tabular}{|c|c|c|c|c|} \hline
    & BER         & BER            & CER        & CER            \\ 
SNR & New method  & Classic method & New method & Classic method \\ \hline
1 & 0.0460 & 0.0640 & 0.20 & 0.45 \\ \hline
2 & 0.0173 & 0.0433 & 0.10 & 0.36 \\ \hline
3 & 0.0153 & 0.0227 & 0.09 & 0.17 \\ \hline
4 & 0.0047 & 0.0080 & 0.03 & 0.07 \\ \hline
5 & 0.0000 & 0.0040 & 0.00 & 0.03 \\ \hline
\end{tabular}
}
\bigskip
\bigskip

Simulation results indicate that the new decoding method
can achieve up to 2--3 dB of coding gain 
on AWGN channel in comparison to classic decoding method.

\section*{Acknowledgment}
The author is grateful to the Alexander von Humboldt Foundation 
and Deutsche Forschungsgemeinschaft (DFG) 
for the many years' support of his research.

\end{document}